\documentclass[fleqn,usenatbib]{mnras}
\usepackage[T1]{fontenc}

\DeclareRobustCommand{\VAN}[3]{#2}
\let\VANthebibliography\thebibliography
\def\thebibliography{\DeclareRobustCommand{\VAN}[3]{##3}\VANthebibliography}

\usepackage{graphicx}
\usepackage{enumerate}
\usepackage{amssymb, amsmath, amsfonts}
\usepackage{gensymb}
\usepackage{mathtools}
\usepackage{natbib}
\usepackage{color}
\usepackage{hyperref}
\usepackage{ulem}
\usepackage{soul}
\usepackage{xspace}
\usepackage{comment}
\usepackage{tabularx}
\usepackage{cellspace}
\newcommand{\Gaia}{{Gaia}\xspace}

\newcommand{\CHARIS}{{CHARIS}\xspace}
\newcommand{\SCExAO}{{SCExAO}\xspace}

\newcommand{\MCFOST}{\texttt{MCFOST}\xspace}
\newcommand{\diskmap}{\texttt{Diskmap}\xspace}

\newcommand{\GPI}{{GPI}\xspace}

\newcommand{\Msun}{\mbox{$M_{\sun}$}}

\newcommand{\Rsun}{\mbox{$R_{\sun}$}}

\newcommand{\Rstar}{\mbox{$R_{\star}$}}
\newcommand{\Mstar}{\mbox{$M_{\star}$}}

\newcommand{\mas}{\hbox{mas}}

\newcommand{\dotdeg}{\rlap{.}^\circ}

\usepackage{newtxtext,newtxmath}

\title[]{Multiband polarimetric imaging of HD 34700 with SCExAO/CHARIS}

\author[M.~Chen et al.]{Minghan Chen$^{1}$\thanks{E-mail: minghan@ucsb.edu},
Kellen Lawson$^{2}$,
Timothy D.~Brandt$^{3}$,
Briley L. Lewis$^{4}$,
Taichi Uyama$^{5}$, 
\newauthor
Max Millar-Blanchaer$^{1}$, 
Ryo Tazaki$^{6, 7}$,
and Thayne Currie$^{8, 9}$
\\
$^{1}$Department of Physics, University of California, Santa Barbara, Santa Barbara, CA 93106, United States\\
$^{2}$NASA-Goddard Space Flight Center, Greenbelt, MD 20771, United States\\
$^{3}$Space Telescope Science Institute, 3700 San Martin Drive, Baltimore, MD, 21218, United States\\
$^{4}$Department of Physics and Astronomy, University of California, Los Angeles, Los Angeles, CA 90095 United States\\
$^{5}$Department of Physics and Astronomy, California State University Northridge,  18111 Nordhoff Street, Northridge, CA 91330, United States\\
$^{6}$Institute of Planetology and Astrophysics, Université Grenoble Alpes, CNRS, F-38000 Grenoble, France\\
$^{7}$Astronomical Institute, Graduate School of Science, Tohoku University, 6-3 Aramaki, Aoba-ku, Sendai 980-8578, Japan\\
$^{8}$ National Astronomical Observatory of Japan, Subaru Telescope, 650 N. Aohoku Pl, Hilo, HI USA 96720\\
$^{9}$ Department of Physics and Astronomy, University of Texas-San Antoino, San Antonio, TX USA 78249
}

\date{Accepted XXX. Received YYY; in original form ZZZ}

\pubyear{2024}

\begin{document}
\label{firstpage}
\pagerange{\pageref{firstpage}--\pageref{lastpage}}
\maketitle

\begin{abstract}
We present Subaru/\SCExAO+ \CHARIS broadband (JHK) integral field spectroscopy of HD 34700 A in polarized light. \CHARIS has the unique ability to obtain polarized integral field images at 22 wavelength channels in broadband, as the incoming light is first split into different polarization states before passing though the lenslet array. We recover the transition disk around HD 34700 A in multiband polarized light in our data. We combine our polarized intensity data with previous total intensity data to examine the scattering profiles, scattering phase functions and polarized fraction of the disk at multiple wavelengths. We also carry out 3D Monte Carlo radiative transfer simulations of the disk using \MCFOST, and make qualitative comparisons between our models and data to constrain dust grain properties. We find that in addition to micron-sized dust grains, a population of sub-micron grains is needed to match the surface brightness in polarized light and polarized fraction. This could indicate the existence of a population of small grains in the disk, or it could be caused by Mie theory simulations using additional small grains to compensate for sub-micron structures of real dust aggregates. We find models that match the polarized fraction of the data but the models do not apply strong constraints on the dust grain type or compositions. We find no models that can match all observed properties of the disk. More detailed modeling using realistic dust aggregates with irregular surfaces and complex structures is required to further constrain the dust properties.

\end{abstract}


\section{Introduction} \label{sec:intro}

Protoplanetary disks are the birthplaces of planetary systems. The initial conditions, stellar radiation, and dynamical interactions between planet formation and disk evolution shape the disk structures and influence the coagulation and growth of dust particles \citep{Tazaki_2019, Benisty_2022}. Observations of disk spectral energy distributions (SEDs) and spatially resolved imaging of disks at various stages of the disk evolution can inform us about the physical processes governing the disk evolution, and help develop self-consistent theories in disk evolution and planet formation \citep{Dullemond_2007, Andrews_2020}. 

Modern high contrast imaging cameras and techniques have become powerful tools to observe circumstellar disks in scattered light. For example, scattered-light observations of PDS 70 with the Very Large Telescope (VLT) in total and polarized intensity resolved the system's disk ring and inner disk and helped to distinguish these signals from those of protoplanets \citep{Keppler_2018, Muller_2018, Mesa_2019, Haffert_2019}; similarly, analysis of Subaru Telescope and VLT observations of AB Aurigae resolved a morphologically complex disk with evidence for an embedded protoplanet(s) \citep[][Dykes et al. 2024, submitted]{Currie2022,Boccaletti2020}.  
The Gemini Planet Imager (GPI) recently observed a large sample of 44 bright Herbig Ae/Be stars and T-Tauri stars, yielding polarized scattered light detections in $80\%$ of them and revealing a wide range of morphologies \citep{Rich_2022}. High contrast imaging in polarized, scattered light has enabled new tests to validate radiative transfer models \citep{Brauer_2019_GGTauri, Arriaga_2020, Bruzzone_2020} and dust aggregate models of disks \citep{Tazaki_2022, Tazaki_2023, Ginski_2023}. 

HD~34700 A is a young spectroscopic binary system (Aa and Ab) with an estimated period of 23.5 days, originally thought to consist of two solar-mass stars and a debris disk, with an age $\gtrsim 10$ Myr \citep{Torres_2004}. However, the \Gaia EDR3 distance places the system at a much further $350.5$~pc \citep{Gaia_2021} compared to the original $\sim$125~pc estimate, and the system was re-evaluated to have an age of $\sim$5~Myr and a transition/protoplanetary disk \citep{Monnier_2019}. Past observations of the disk include near-infrared (J and H) broadband total intensity and polarized intensity by the Gemini Planet Imager (\GPI) \cite{Monnier_2019}, millimeter images by the Submillimeter Array \citep{Benac_2020}, and integral field spectroscopy images in total intensity in JHK by the Coronagraphic High Angular Resolution Imaging Spectrograph (\CHARIS) coupled with the Subaru Coronagraphic Extreme Adaptive Optics (\SCExAO) \citep{Uyama_2020}. These observations have revealed that the disk around HD~34700 A is rich with sub-structures: numerous spiral arms, a cavity at the north, and many darkening features. Most recently, the Spectro-Polarimetric High-contrast Exoplanet REsearch facility (SPHERE; \citealt{Beuzit_2019}) at the VLT was able to fully resolve the inner disk around HD 34700 at a separation of $\sim$65~au to $\sim$117~au \citep{Columba_2024}.

In this paper, we present new \SCExAO+ \CHARIS integral field spectroscopic observations of the disk around HD~34700 A in polarized light. In Section \ref{sec: data}, we describe our unique spectropolarimetric dataset and the data reduction processes. In Section \ref{sec: imaging analysis}, we present the reduced images of the disk, describe the disk morphology and features, and measure the surface brightness profiles and phase functions of the disk. In Section \ref{sec: modeling} we carry out 3D radiative transfer modeling of the disk at multiple wavelengths in JHK, and discuss our results and their implications. Finally, we summarize our work in Section \ref{sec: discussion}.

\section{Observation and Data Reduction} \label{sec: data}
\subsection{CHARIS Observation Data} \label{subsec: observation}
We present the polarimetric imaging results of HD 34700 A at J, H and K bands from \CHARIS \citep{CHARIS_Design, Groff_2014_CHARIS_contruction} coupled with \SCExAO \citep{AO188, Jovanovic_2015_SCExAO}. We also use an earlier total intensity dataset presented and analyzed by \cite{Uyama_2020} for comparison and to calculate polarized fraction. The stellar properties of the HD 34700 A system are listed in Table \ref{table: stellar properties}.

\begin{table}

\caption{Stellar Properties}
\label{table: stellar properties}
\begin{tabularx}{0.48\textwidth}{>{\centering}p{2cm}>{\centering}p{4cm}c}
\hline
Property & Value & Reference\\
\hline
R.A. (J2000) & 05:19:41.41 & 1\\
DEC. (J2000) & +05:38:42.80 & 1\\
Distance & $350.5 \pm 2.5$ pc & 1\\
Binary Period & 23.49 days& 2\\
Primary $T_{\text{eff}}$ & 5900 K & 2\\
Secondary $T_{\text{eff}}$ & 5800 K & 2\\
Spectral types & G0 IVe, G0 IVe& 3\\
J mag & $8.041 \pm 0.023$ & 4 \\
H mag & $7.706 \pm 0.023$ & 4 \\
Ks mag & $7.482 \pm 0.024$ & 4 \\
Age & $\sim 5$ Myr & 5\\
$M_1$, $M_2$ & $2.05 \Msun$, $2.04 \Msun$ & 5\\
\hline

\end{tabularx}
\textbf{References:} 1. \cite{Gaia_2021}; 2. \cite{Torres_2004}; 3. \cite{Mora_2001}; 4. \cite{Skrutskie_2006}; 5. \cite{Monnier_2019};

\end{table}

\subsubsection{Total Intensity Data} \label{subsec: ti data}

We use the total intensity data of HD~34700~A taken by Subaru/\SCExAO $+$ \CHARIS on 2019 January 12 UT with a fixed pupil and a Lyot coronagraph mask to suppress starlight (Principal Investigator [PI]: Thayne Currie). The data were taken in broadband integral field spectroscopy (IFS) mode. In this mode, CHARIS covers a wavelength range of $1.16-2.37~\mu {\rm m}$ at a spectral resolution of $R \sim 19$ with a pixel scale of $16.15\; \mas\; \text{pixel}^{-1}$ \citep{Currie2023, Chen_2023}. The total intensity data was previously analyzed and published by \cite{Uyama_2020}. 
HR 2466 was also observed for a PSF reference for reference-star differential imaging (RDI; \citealt{Lafrenière_2009_RDI}). The data were taken under very good seeing conditions ($\theta_{V}\sim 400~\mas$); the typical Full Width at Half Maximum (FWHM) was $\sim 30, 45, 55\; \mas$ in JHK bands, respectively. The total exposure time was 2168.6 seconds ($30.98$-second exposures $\times$ 70 cubes) for HD 34700 A and 2952.95 seconds (1.475-second single exposure $\times$ 14 coadds $\times$ 143 cubes) for HR 2466. We re-reduce the total intensity data for this work using an improved RDI PSF subtraction algorithm described in Section \ref{subsec: RDI PDI}.

\subsubsection{Polarized Intensity Data} \label{subsec: pi data}

We use two sets of polarized intensity data taken on two different dates. We will call them the primary polarized intensity data and the reference polarized intensity data. 

Our primary data were taken on UT December 16 2019 (PI: Tim Brandt) with \CHARIS broadband in polarimetry mode. In this mode, the incoming light is separated into two orthogonal polarization states by a Wollaston prism, and split into two beams by a beamsplitter, before entering the \CHARIS lenslet array.  The light then passes through a prism to be dispersed into a grid of microspectra on the detector. As a result, two $1''\times2''$ sections corresponding to the two orthogonal polarization states of the same field of view are dispersed onto the \CHARIS detector, which originally covers a field of view of $2''\times2''$. As the light is first separated into two polarization states and then passed through the integral field spectrograph, we can extract Stokes parameter images at the full spectral resolution of \CHARIS broadband: 22 wavelengths spanning JHK for every spatial element in the $1''\times2''$ field of view. This is a unique advantage to CHARIS data, as most other polarimetry instruments can only observe at broad filters (e.g. \GPI; \citealt{Perrin_2015, Monnier_2019}) SPHERE; \citealt{de_Boer_2020}) or at a specific wavelength (e.g. ZIMPOL; \citealt{Columba_2024}).

The \CHARIS polarimetry mode allows us to obtain full-frame polarized spectral data for HD 34700. The integration time for each exposure was 50.16 seconds for a total of 146 exposures ($\sim$122 minutes). A half-waveplate was inserted and cycled through angles $0 \degree$, $45 \degree$, $22\dotdeg 5$ and $67\dotdeg 5 $, with two exposures at each angle in every cycle. A Lyot coronagraph was used to block out the central binary. 

In addition to our primary data, we use a reference CHARIS PDI dataset taken in the same polarimetry mode on UT 2019 Feb 26 (PI: Thayne Currie), but with the standard 4-satellite-spot pattern induced by SCExAO for astrometric and spectrophotometric calibration. The reference dataset has 284 exposures and an integration time of 7 seconds for each exposure. We use this reference dataset for flux calibration purposes to correct for systematics potentially caused by the ``alternating grid'' calibration pattern of our primary data (described in Section \ref{subsec:centroiding}). We describe the systematics and the flux calibration in Section \ref{subsec: fluxcal}. The standard reduction and calibration of both PDI datasets follow the procedure described in Section \ref{subsec: RDI PDI}. Then, our primary polarized intensity data is further corrected for systematics based on the reference polarized intensity data. The reference data has much less total integration time than the primary data, and covers a parallactic rotation of only $\sim 3\degree$. Without enough field rotation, the reference data only imaged the disk partially due to the small field of view. Therefore, we use the reference data only for calibration purposes. The calibration procedure is described in detail in Section \ref{subsec: fluxcal}. 

\subsubsection{Polarized Flat Field Data} \label{subsec: pi flatfield data}
After cube extraction, the integral field spectropolarimetry data were adjusted with additional flat-field images, termed ``PDI flat-fields," to correct effects introduced by the extra polarimetry optics (which are not included in the optical path for the standard CHARIS flat fields). PDI flat-field images for CHARIS are collected by observing SCExAO's halogen lamp calibration source \citep{Jovanovic_2016} with the Wollaston prism in place. For this purpose, the PDI flat-field images are extracted into cubes exactly as for the science data. The flat-field cubes are then median combined to create a single flat-field image cube. This cube is then normalized by dividing each wavelength slice by the median value across the slice. Subsequently, each science cube is flat-fielded by dividing it by this master PDI flat-field cube. Compared to the distributed flat fields in the CHARIS Data Reduction Pipeline (CHARIS DRP) \citep{Brandt_2017_CHARISPipeline}, these polarized flat fields show different structures for the two $1''\times2''$ field of views, and have variations in transmission of up to $10 \%$ which would bias the final Stokes parameters if not corrected.

As no PDI flat-fields were taken contemporaneously with the data, the nearest available calibration data from UT 2021 March 20 were adopted for this purpose. Comparison with more recent PDI flat-field images shows only small changes over time, such that the impact of time difference is expected to be negligible. 

\subsection{Data Reduction} \label{subsec:reduction}
\subsubsection{Cube Extraction and Selection Criteria}
The raw data were first extracted into data cubes using the CHARIS DRP \citep{Brandt_2017_CHARISPipeline}. Each data cube consists of 2D images at 22 wavelength channels that span the range of J, H and K bands. The data cubes are spectrophotometrically calibrated to the bright primary stars using Kurucz model atmospheres \citep{Castelli_Kurucz_2004}, adopting spectral type G0V for HD 34700 A and A2V for HR 2466 \cite{Cutri_2003}. For the polarized intensity data, we experienced technical issues with optics alignment, a period of time with bad seeing, and varying adaptive optics (AO) system performance during the observing sequence. We visually inspect each exposure and manually remove ones with poor AO correction or seeing, and ones with instrumental artifacts. As a result, 70 exposures were discarded and 76 exposures were kept for the analysis. 

\subsubsection{Image Registration Using Alternating Satellite Spots}
\label{subsec:centroiding}

To locate the centroid of the central star, align the images in coronagraphic data, and calibrate the flux of the disk's surface brightness, a diffractive grid generated by sine waves on the deformable mirror (DM) of \SCExAO in the pupil plane projects the primary star onto the focal plane with a fixed contrast at fixed separations and angles \citep{Jovanovic_2015, Jovanovic_2015_SCExAO, Sahoo_2020}. Usually, there are four satellite spots at a fixed $\lambda / d$ separation that are roughly at the vertices of a square centered at the central star. However, the satellite spots contaminate a significant area of the disk in this case. To mitigate this contamination, we adopted an alternating pattern of satellite spots \citep{Sahoo_2020}, where only two spots along one diagonal are projected in each exposure, and it alternates between the two diagonals over the sequence. This allows all parts of the disk to be free of satellite spots and therefore uncontaminated for half of the exposures, potentially providing better recovery of the disk signal, as well as better isolation of the satellite spots from the disk signal for a better flux calibration. The satellite spots are fitted and used to triangulate the centroid of the primary star in order to register the images. This is done using a centroiding algorithm described in \cite{Chen_2023}. The re-registered images are then used for PDI reduction.

\subsubsection{RDI and PDI} \label{subsec: RDI PDI}
Once registered, the PDI data cubes are further reduced following a similar procedure as that described in \citet{Lawson2021}, but with the updates and caveats summarized hereafter. Initial spectrophotometric calibration is conducted following \citet{Lawson2021}, but using only the cubes with visible satellite spots in every channel. Because the original field of view is split into two halves, the satellite spots in one of the two alternating diagonals move beyond the field of view after the seventh wavelength channel. Therefore, we only measure the fluxes of the satellite spots along the diagonal that stay within the field of view, and adopt this calibration for the exposures with the other configuration. The remaining cubes adopt the flux calibration scaling factors for the nearest calibrated cube in time. Results for this spectrophotometric calibration strategy were comparable to those achieved when first subtracting the alternating cubes to better isolate the flux of the satellite spots. Matching of half-waveplate cycles and the PDI procedure itself are carried out following the procedure outlined in \citet{Lawson2021}. In place of the first order correction for instrumental polarization described therein, instrumental polarization is corrected using the full Mueller Matrix model summarized in \citet{Hart2021}.

\begin{figure}
    \centering
    \includegraphics[width=0.48\textwidth]{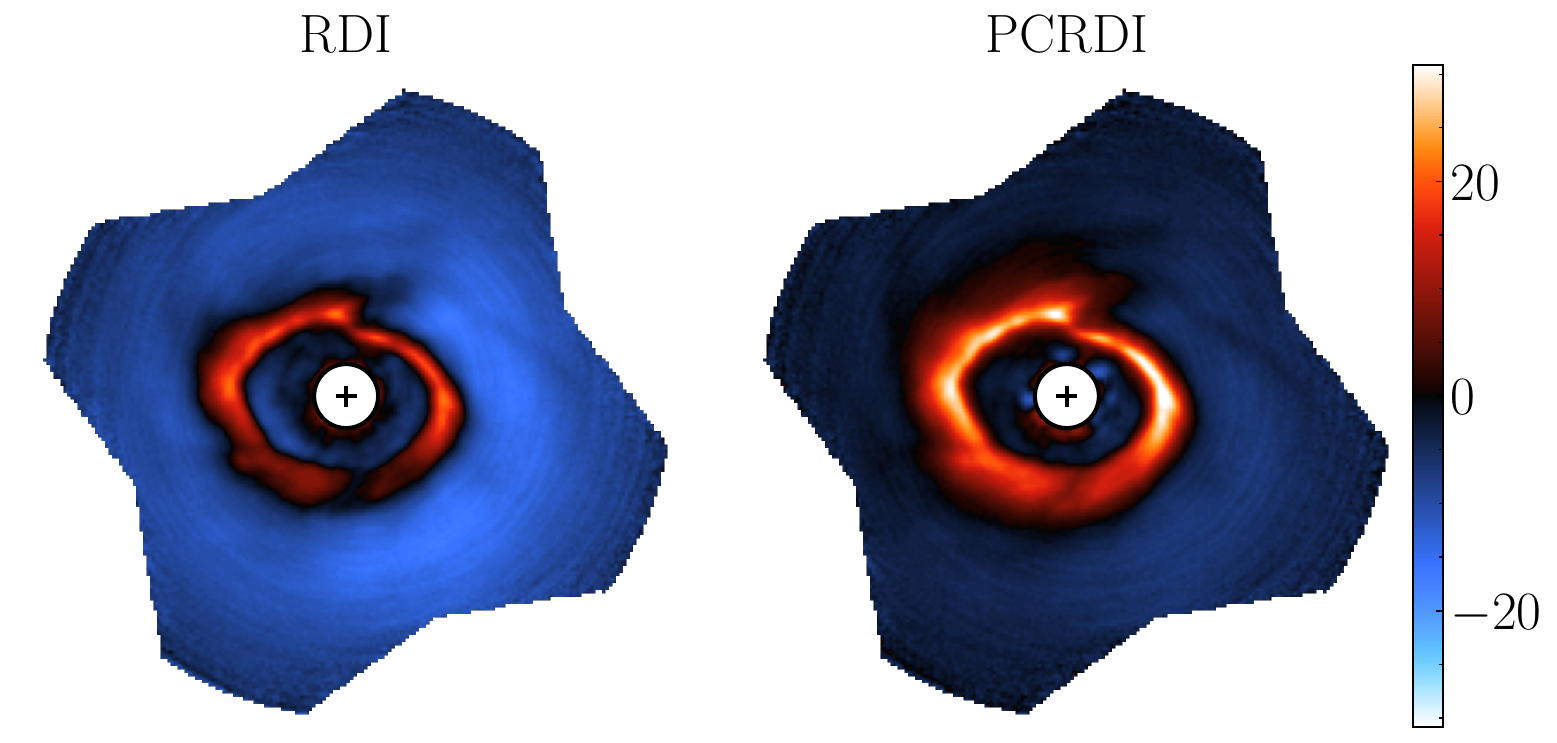}
    \caption{A comparison of the standard RDI with polarimetry-constrained RDI (PCRDI). The flux is in arbitrary units, scaled by the deprojected radius of the disk surface for better visual clarity. PCRDI is shown to visibly improve the over-subtraction of the disk signal and of the background.}
    \label{fig: pcrdi}
\end{figure}

To carry out PSF subtraction of the total intensity data from \citet{Uyama_2020}, we used the polarimetry-constrained reference star differential imaging (PCRDI) technique \citep{Lawson2022}. In PCRDI, available polarized intensity images are used to estimate the total intensity of the circumstellar signal (CSS) in order to suppress its impact on the creation of a stellar PSF model — and thus to eliminate the over-subtraction that typically plagues PSF-subtracted total intensity products. Our procedure for optimizing the CSS estimate in PCRDI diverges slightly from the one introduced in \citet{Lawson2022}. Rather than optimizing a function of five parameters — four governing the scattering surface and an overall scaling factor — only the four parameters of the scattering surface are allowed to vary. For each trial solution, the resulting polarized intensity-based CSS estimate is treated like a disk model in RDI forward modeling and then compared with the residuals for a standard (unconstrained) RDI reduction for the real data. This allows the optimal scaling factor for each wavelength channel to be determined analytically by minimizing the sum of the squared residuals between the (scaled) forward modeled image and the data. Once the optimal parameters are identified, the corresponding CSS estimate is used to carry out PCRDI for the total intensity data. A comparison of PCRDI with the standard RDI is shown in Figure \ref{fig: pcrdi}. The flux is scaled by the de-projected radius of the disk surface to make the difference more distinguishable visually. In PCRDI, the disk is visibly brighter and suffers less over-subtraction. The surrounding area is also much closer to zero, compared to over-subtracted negative areas in standard RDI. 

\subsubsection{Polarized Intensity Calibration} \label{subsec: fluxcal}
For coronagraphic data, the brightness of the astrophysical source is usually calibrated using the satellite spots and the known contrast of the spots to the central host star. Then, using the known magnitude of the host star we can convert the data units to physical units for the astrophysical source, which is the disk in this case. Since we are observing an extended disk signal rather than a point source, a fixed four-spot pattern would permanently overlay parts of the disk, potentially introducing biases the current data reduction pipeline is not equipped to correct for. Therefore, we opted to use an alternating two-spot pattern described in Section \ref{subsec:centroiding}, which leaves all parts of the disk unobscured for at least half the exposures. The contrast between the satellite spots and the host star for SCExAO is calibrated using an internal source over a narrow bandpass. This contrast measured at a specific DM amplitude and wavelength, and scales as $A^2 \lambda^{-2}$, where $A$ is the modulation amplitude of the DMs, and $\lambda$ is wavelength. The spot contrasts were measured to a precision of a few percent for both the regular four-spot pattern as well as the alternating two-spot pattern, and the grid amplitude and wavelength dependence are found to agree with the expected scaling \citep{Currie_2020SPIE, Sahoo_2020}. However, by adopting the calibration results in \cite{Sahoo_2020}, we find the calibrated surface brightness of the disk of our primary polarized intensity data to be systematically fainter by about a factor of 3 fainter than previously published polarized intensities in J and H bands \cite{Monnier_2019}, and also fainter by a similar factor than the reference polarized intensity data, described in Section \ref{subsec: pi data}.

This systematic difference in the flux calibration is likely due to the unconventional alternating two-spot pattern. For the fixed four-spot pattern, the spot contrasts remain relatively stable over time between different tests. The stability of this mode has been quantified in \citet{Currie_2020SPIE}, where an $\sim 8 \%$ variation in the contrast was found between two calibration tests at different times. However, for the alternating two-spot pattern, no such investigation on contrast stability over different epochs has been carried out to our knowledge. We therefore assume the combined effect of hardware and software handling of the rarely used two-spot pattern is likely responsible for this large systematic shift in our flux calibration. To address this shift, we use the reference polarized intensity data described in Section \ref{subsec: pi data} to fit for a scaling factor. We use the mean-collapsed images of J and H bands of both datasets for calibration, as K band images have $\sim 5-10$ times lower signal-to-noise ratios (SNR). For each collapsed image, we find the best-fit global scaling factor that scales our primary data to the reference data in polarized intensity. When fitting for this scaling factor, we consider a 10-pixel wide annulus centered at the best-fit ellipse to the disk's peak brightness, described in Section \ref{subsec: ring}). To estimate the mean and error of the scaling factor, we bootstrap resample the wavelengths in J and H and create data cubes consisting of images at the bootstrapped wavelengths in J and H, then collapse the these data cubes and fit for the scaling factors for these bootstrapped samples. We find an overall calibration factor of $3.18 \pm 0.06$. This factor scales our polarized intensity data to the reference \CHARIS data described in Section \ref{subsec: pi data}, observed in the same polarimetry mode, but with a diffractive grid configuration that has a robust flux calibration. The uncertainty represents only the scatter of the scaling factor among the bootstrapped samples, and is added to the other uncertainties of the flux calibration in quadrature. A small uncertainty here reassures that this systematic is not wavelength dependent.

\section{Imaging Analysis} \label{sec: imaging analysis}

\begin{figure*}
    \centering
    \includegraphics[width=0.95\textwidth]{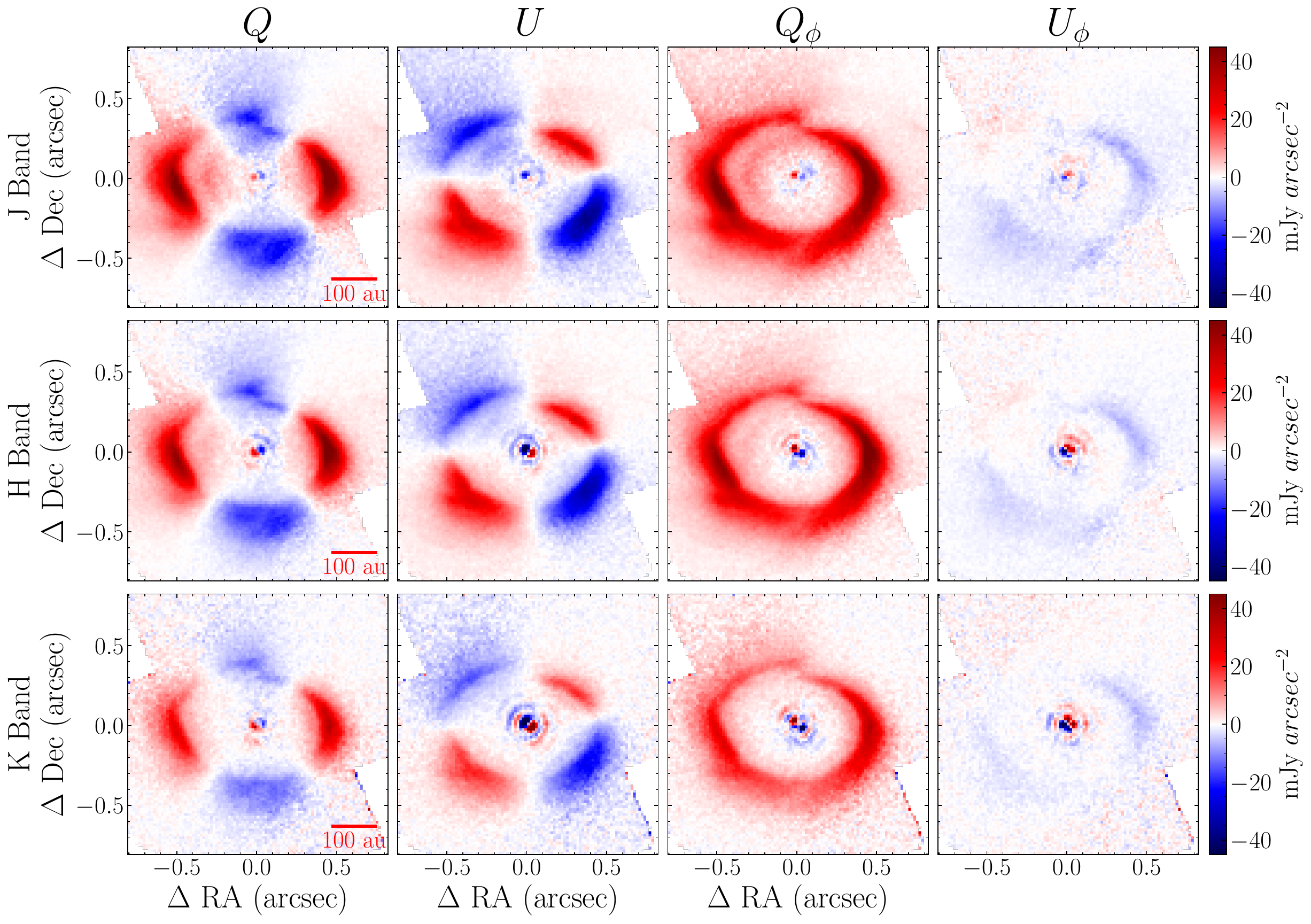}
    \caption{The Stokes Q (first column), U (second column) and radial Stokes $Q_{\phi}$ (third column), $U_{\phi}$ (fourth column) images of the disk, collapsed into J (top row), H (middle row), and K bands (bottom row). Astrophysical signal is present in $U_{\phi}$ with an average SNR of around 4 to 5 depending on wavelength, and up to an SNR of 10 at the brightest regions in J band.}
    \label{fig: StokesJHK}
\end{figure*}

The double-differencing procedure described in \cite{Lawson2021} gives Q and U Stokes parameter images at each wavelength. These are collapsed in to J, H, and K broadband images shown in Figure \ref{fig: StokesJHK}. The first two columns show the typical quadrant-patterned Q and U Stokes images in J, H and K. The last column shows the radial Stokes images, $Q_{\phi}$. Radial Stokes parameters, $Q_{\phi}$ and $U_{\phi}$, are defined as:
\begin{align}
    \label{eqn:Qphi}
    Q_{\phi} = -Q \cos2\phi - U \sin2\phi\\
    \label{eqn:Uphi}
    U_{\phi} = +Q \sin2\phi - U \cos2\phi
\end{align}
Radial Stokes parameters are first introduced in \cite{Schmid_2006} as $Q_{r}$ and $U_{r}$, which are projections of Q and U onto the azimuthal and radial directions ($\pm Q_{\phi}$), and $45 \degree$ angles from those directions ($\pm U_{\phi}$). The radial Stokes parameters are useful for two main reasons: 1. In the case of single scattering events, the polarization would be entirely in the azimuthal direction and appear as $+Q_{\phi}$, while $U_{\phi}$ provides an estimate of the noise \citep{Monnier_2019}; and 2. For noisy data, the polarized intensity 
\begin{align}
    \label{eqn:pol flux}
    I_{p} = (Q^2 + U^2)^{1/2}
\end{align}
will have a positive bias, while $Q_{\phi}$ does not. The noise in Q and U, originally centered around zero for a well-calibrated dataset, would now become strictly positive after taking the sum of the squares \citep{Schmid_2006}.

These properties make $Q_{\phi}$ a good alternative for the total polarized flux in Equation \ref{eqn:pol flux}, and $U_{\phi}$ a good proxy for the noise when analyzing disk signals. However, large optical depths, inclination, asymmetric scattering combined with PSF convolution can all influence the polarization and produce signals in $U_{\phi}$ in many situations \citep{Monnier_2019, Canovas_2015, Dong_2016}. For HD~34700, \cite{Monnier_2019} found significant signals in $U_{\phi}$, which we also found in our \CHARIS polarized dataset. The average SNR of the disk signal per pixel area ($\sim 261 \; \mas^2$) in $U_{\phi}$ is around 3-5 in the collapsed J, H, K images, with the brightest region in J band reaching a maximum SNR of around 10. Therefore, we do not use $U_{\phi}$ as a proxy for the noise estimate, and instead use a more typical standard deviation approach described in Section \ref{subsec: SB profiles}. Furthermore, $Q_{\phi}$ is no longer equivalent to total polarized light, $I_p$, for this disk; however, it remains a good alternative to $I_p$, provided that we also project our models into $Q_{\phi}$. Whether our signal in $U_{\phi}$ results from real dust scattering physics or from geometric/PSF convolution effects requires careful analysis and modeling of the $U_{\phi}$ component of the disk, which is beyond the scope of this paper. We restrict our analysis and modeling to the $Q_{\phi}$ component only. For simplicity, for the remainder of this paper, we will use ``$Q_{\phi}$'' and ``polarized intensity'' interchangeably, and refer to ``$Q_{\phi}/I$'' as the polarized fraction, where $I$ is the total intensity. We note that we adopt the radial Stokes conventions in Appendix A in \cite{Monnier_2019}, where polarization along the azimuthal direction corresponds to a positive signal in $Q_{\phi}$.

\begin{figure*}
    \centering
    \includegraphics[width=0.95\textwidth]{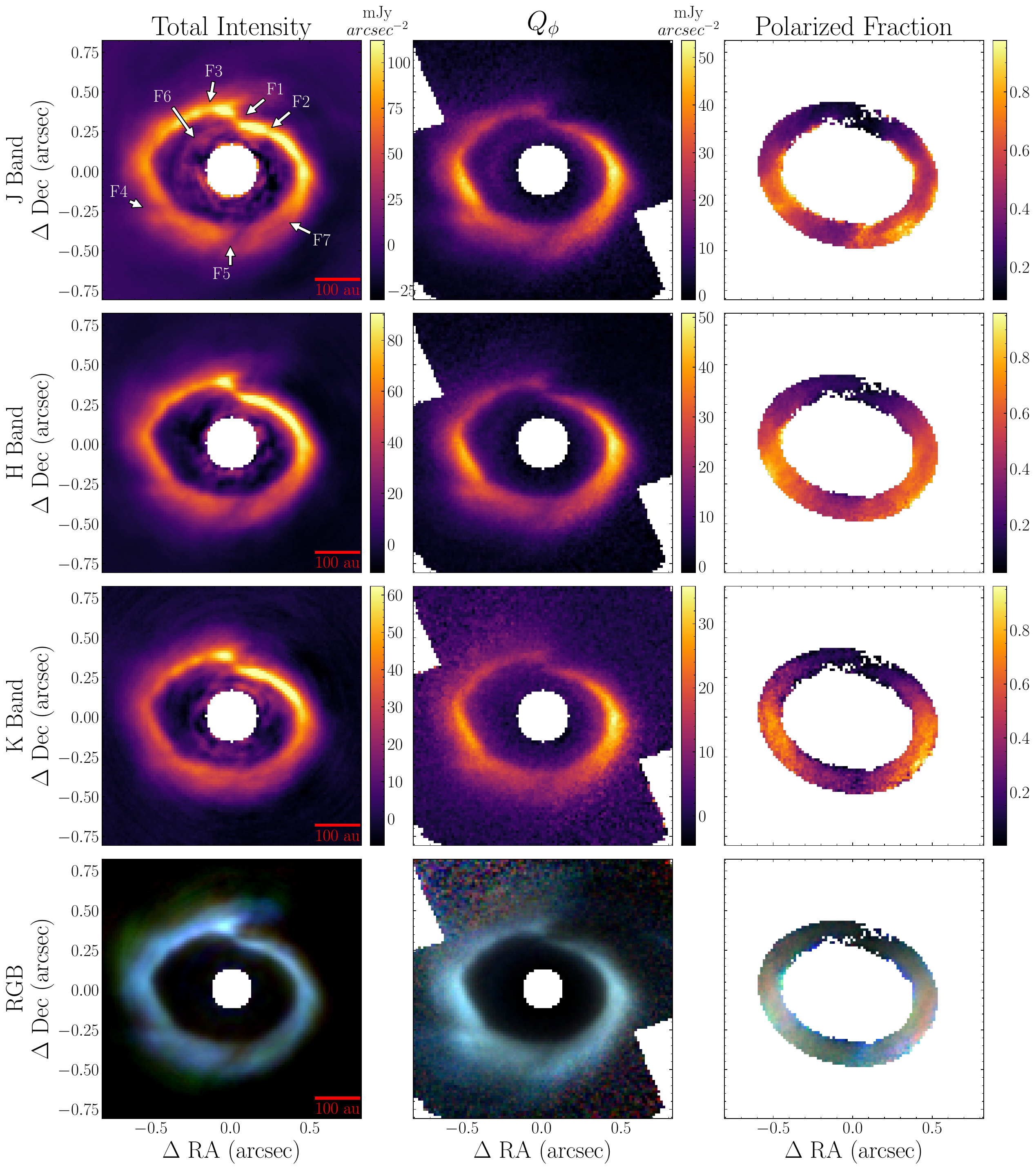}
    \caption{First three rows: collapsed J, H, K band images. Left column: total intensity from \protect\cite{Uyama_2020}, re-reduced with PCRDI that minimizes over-subtraction. Middle column: $Q_{\phi}$ of our polarized intensity data. Right column: polarized fraction, $Q_{\phi}/I$, where $I$ is the total intensity, in the bright disk ring region. The bottom row contains false color composite renderings of the images as RGB images (R=K, G=H, B=J), with the maximum flux in J band normalized to the maximum value of blue. The various disk features are annotated in the very first panel, including the cavity (F1), darkening features (F2-F5), an inner arc F6, and an instrumental artifact (F7)}
    \label{fig: RGB}
\end{figure*}

Our polarized intensity images obtained simultaneously at 22 wavelengths from $1.16~\mu {\rm m}$ to $2.37~\mu {\rm m}$, combined with the total intensity data from \cite{Uyama_2020} taken in the same CHARIS broadband mode, allow us to examine the wavelength dependence of the polarization properties in more detail than was previously possible. We show the total and polarized intensity images collapsed into J, H, and K band in Figure \ref{fig: RGB}, as well as the polarized fraction images obtained by taking the ratio between corresponding polarized and total intensity images. We also show false color composite renderings using these images, with J, H, and K images acting as the blue, green and red color channels in an RGB image, respectively.

To reduce the cost of computation in disk modeling and at the same time boost the SNR without losing too much resolution in wavelength, we binned our wavelength channels into slightly larger wavelength bins. First, we discard the two wavelength bins at $1.37~\mu{\rm m}$ and $1.87~\mu {\rm m}$ in which atmospheric transmission is low, and also the two reddest wavelengths at $2.29~\mu{\rm m}$ and $2.37~\mu{\rm m}$ which have SNR$<3$ for most areas of the disk. We bin the remaining 18 wavelengths by groups of three to obtain 6 wavelength bins centered at $1.2, 1.3, 1.5, 1.7, 1.9,$ and $ 2.1~\mu{\rm m}$. The data are mean-collapsed within each bin. Subsequent analysis and modeling are all carried out at these wavelengths.

\subsection{Morphology of The Dominant Ring} \label{subsec: ring}

The disk around HD~34700~A is dominated in scattered light by a bright ring $\sim$0$.\!
\!''5$ in radius and rich with structure.  \cite{Monnier_2019} and \cite{Uyama_2020} both characterized this ring; \cite{Monnier_2019} used both total and polarized intensity, while \cite{Uyama_2020} used only total intensity.  Both analyses involved fitting an ellipse to the ring. They obtained similar inclinations and inner ring radii, but slightly different position angles (PA) of the semi-major axis: $69\dotdeg 0 \pm 2\dotdeg 3$ \citep{Monnier_2019} and $74\dotdeg 5 \pm 1\dotdeg 0$ \citep{Uyama_2020}. This difference ($\sim 2\sigma$) is modest considering that the datasets are taken by different instruments and the ellipses are fit using different methods. In this section we revisit the basic morphology and orientation of the ring using our CHARIS data. 

We adopt three approaches to fit for the position and morphology of the dust ring.  For our first fitting method (abbreviated ``lsq distance'' in Table \ref{table:ellipse fit}), we take a very similar approach to that used in \cite{Monnier_2019} and \cite{Uyama_2020}. We divide the ring into small azimuthal sections and compile a list of the pixel coordinates of the local radial maxima for each section. We use the Python least-squares ellipse fitting software given by \cite{ben_hammel_2020}, the same software used by \cite{Uyama_2020}. It performs a least-squares fit by minimizing the sum of the squares of the algebraic distances of the provided data points to the ellipse, and is shown to be robust against noise. Our second and third fitting methods maximize the sum of the fluxes of the pixels the ellipse passes through instead of minimizing the least-squares distance of a collection of pixels. In our second method (``max flux''), we take the average intensity of all the pixels a trial ellipse passes through, and maximize this average. The third method (``max flux interp'') takes a set of evenly spaced (in terms of arclength) points on the ellipse, and interpolates the values at the exact coordinates of those points. The second and third methods maximize similar metrics, and thus produced similar results as expected ($<1\sigma$). They agree with the first method to $\sim 2\sigma$.

There are two major features of the disk that produce notable deviations from an elliptical ring. One is the discontinuity on the north side of the disk; the other is an arc-like feature on the south side of the disk that is fainter in polarized intensity than its surroundings (an under-dense spiral arc). These features are visible in Figure \ref{fig: RGB}, labeled F1 and F5 respectively. When selecting the local radial maxima around the ring for the first ellipse fitting method, the maxima at these features clearly deviate from the ellipse. Additionally, the flux changes at these features could also potentially bias the other two fitting methods that use flux maximization. Therefore, for all three methods, we mask out a $60\degree$ section at the north side and a $30\degree$ region at the south side. The north side mask covers the large cavity (F1) and a darkening feature (F3). The south side mask covers another darkening feature F5 features. Compared to the fits without the mask, the best-fit inclination and PA change between $0-2\sigma$. The errors of the best-fit parameters for the first method are calculated via bootstrap resampling of the points selected for fitting. For the second and third methods, while we mask out two major disk features when fitting, the many arcs of the disk could have small effects on the radial profile that are not obvious by eye. Therefore, to estimate the uncertainties, we divide the remaining unmasked disk into 90 azimuthal sections with 3 degrees per section. We then carry out jackknife resampling by masking out one section at a time and fitting for the ellipse for each Jackknife sample, and estimate the uncertainties using these fits. 

To account for systematic differences between these fitting methods, we take the error-weighted average of the three methods to derive our final reported disk parameters, summarized in Table \ref{table:ellipse fit}. We also include the uncertainty of the CHARIS instrument's north pointing \citep{Chen_2023} in our errors on the inferred PA, and add all uncertainties in quadrature for our final error. The best-fit inclination and semi-major axis are in good agreement ($< 1\sigma$) with the best-fit values in \cite{Monnier_2019} and \cite{Uyama_2020}. The best-fit PA, $74\dotdeg 3 \pm 1\dotdeg 2$, agrees with that in \cite{Uyama_2020} within $1\sigma$, but differs by $\sim 2\sigma$ from the best-fit value in \cite{Monnier_2019}, $69\dotdeg 0 \pm 2\dotdeg 3$.

\begin{table*}

\caption{Best-fit ellipse parameters for different fitting methods}
\label{table:ellipse fit}
\begin{tabular}{ccccc}
\hline
Fit method$^{1}$ & Semi-major axis (\mas) & Inferred Inclination & Inferred PA (\degree) & Displacement from star\\
 & & & & (separation (\mas), PA (\degree))\\
\hline
lsq distance & $488.4 \pm 2.7$ & $39.7 \pm 0.7$ & $73.1 \pm 0.9$ & (52.1, 129.0) \\
max flux & $494.0 \pm 2.5$ & $41.6 \pm 0.5$ & $75.0 \pm 0.6$ & (43.5, 135.5) \\
max flux interp & $492.8 \pm 3.1$ & $41.5 \pm 1.6$ & $75.0 \pm 1.8$& (47.5, 130.4) \\
\hline
\multicolumn{5}{c}{Error-Weighted Average + Errors}\\
 & $491.8 \pm 3.3$ & $41.0 \pm 1.1$ & $74.3 \pm 1.2$ & \\
\hline
\end{tabular}\\
\end{table*}

\begin{figure*}
    \centering
    \includegraphics[width=\textwidth]{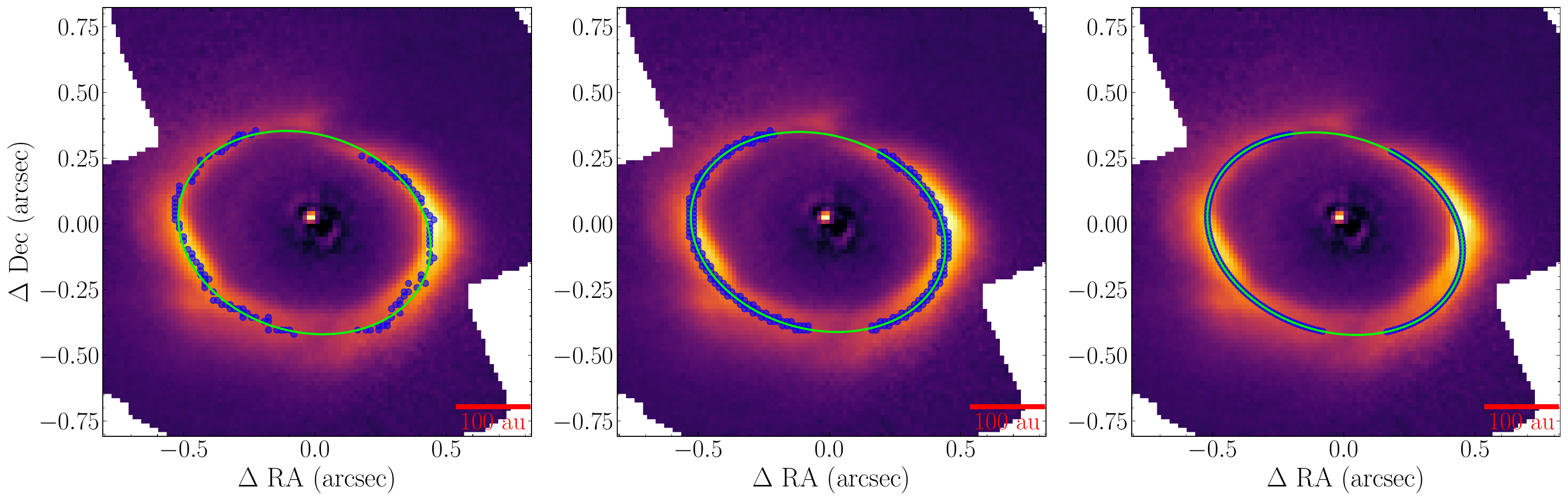}
    \caption{Ellipse fitting to the disk. The three panels correspond to three different fitting methods: least-squares distance (left), maximum flux (middle) and interpolated maximum flux (right). In the first two cases, the blue points are the pixels selected to calculate distance or flux. For the last case, the blue points are the equally spaced points on the ellipse for interpolation. The green lines are the best-fit ellipses to the blue points in each case. The best-fit values for semi-major axis, inclination and PA agree within $1\sigma$ between the second and third methods, and they agree with the those of the first method to $\sim 2\sigma$; these values are reported in Table \ref{table:ellipse fit}.}
    \label{fig:ellipse fit}
\end{figure*}

\subsection{Disk Features}\label{subsec: disk features}
In Figure \ref{fig: RGB} we annotate the various features of the disk, including the cavity at the north side of the disk (F1), the darkening features (F2-F5), an arc that is part of the inner disk (F6), and an instrumental/reduction artifact (F7). The features labeled F1-F6 have been discussed in detail in past works \cite{Monnier_2019, Uyama_2020, Benac_2020} and we do not focus on studying them in more detail in this paper.  F6 shows a faint inner arc the J band images, which is more clearly seen in \cite{Monnier_2019} and is confirmed to be part of a mis-aligned inner disk \citep{Columba_2024}. The inner disk could cast shadows onto the outer disk, changing the intensity and scattering profiles \cite{Brauer_2019_GGTauri}. This poses major difficulties in the radiative transfer modeling and subsequent interpretation of the disk, which we discuss in more detail in Section \ref{sec: modeling}. We observe that the wavelength dependence of the brightness of most of these features remain consistent with the rest of the disk and appear blue in the RGB image, suggesting broad consistency in the dust properties. The exceptions are the darkening features F4 and F5, at PA$\sim 120\degree$ and $\sim 180\degree$, respectively. In these regions, the wavelength dependence is noticeably shallower than the rest of the disk. F6 is an arc that is part of the inner disk, which is both too faint and too close to the inner working angle to be fully resolved by CHARIS. Finally, F7 is an arc-like feature that can be seen in the J band. It is more visible in the J band polarized fraction panel. In an un-collapsed cube with five wavelength channels in J band, this feature can be seen radially expanding through wavelengths as would be expected for a diffraction feature originating from the star.  We consider F7 to most likely be an instrumental/data reduction artifact.

\subsection{Surface Brightness Profiles}
\label{subsec: SB profiles}
To measure surface brightness profiles of the disk, we first convolve the collapsed images in J, H, and K with a normalized Gaussian kernel with $\sigma=1.0$ pixels. The Gaussian kernel is smaller than the size of the instrumental PSF, and thus can reduce the noise and preserve the disk structures. This effectively measures the local surface brightness. We measure the surface brightness along the azimuthal direction. We take all the pixels intersected by the best-fit ellipse and measure the surface brightness and PAs of these pixels, where the PA of a pixel refers to the PA of the position of the pixel relative to the central star. The polarized fraction is defined as $Q_{\phi}/I$, where $I$ is the total intensity. 

To measure the errors of each data point, we first estimate a corresponding noise image for every collapsed image. We subtract from each image the median filtered image with a filter size of 3 pixels, which provides an estimate for the noise assuming there is negligible structure on the scale of the size of the filter. We divide this noise image into 11 concentric annuli around the central binary, from a radius of 10 pixels to a radius of 65, roughly corresponding to the radii of the inner working angle and the extent of the field of view. Thus, each annulus has a width of 5 pixels. We measure the standard deviation of the residuals for each annulus and use that as the noise estimate at the radius of the bin center. This gives us radially dependent noise estimates at 11 radii. Finally, we interpolate the noise at each pixel's radius to obtain the noise image. We then add the uncertainties of the flux calibration described in Section \ref{subsec: fluxcal} to this noise image in quadrature. To properly propagate the noise for the convolved intensity scattering profiles, the squares of the noise images (i.e. variance images) are convolved with the squares of the Gaussian kernels. The noise image is about a factor of 2-3 times higher than the flux calibration uncertainty for the bright part of the disk, but they are on the same order of magnitude, $\sim 2\%-5 \%$ relative to the bright part of the disk.  This gives a typical SNR of $\sim$20-50 per pixel area in the bright parts of the disk. 

\begin{figure}
    \centering
    \includegraphics[width=0.48\textwidth]{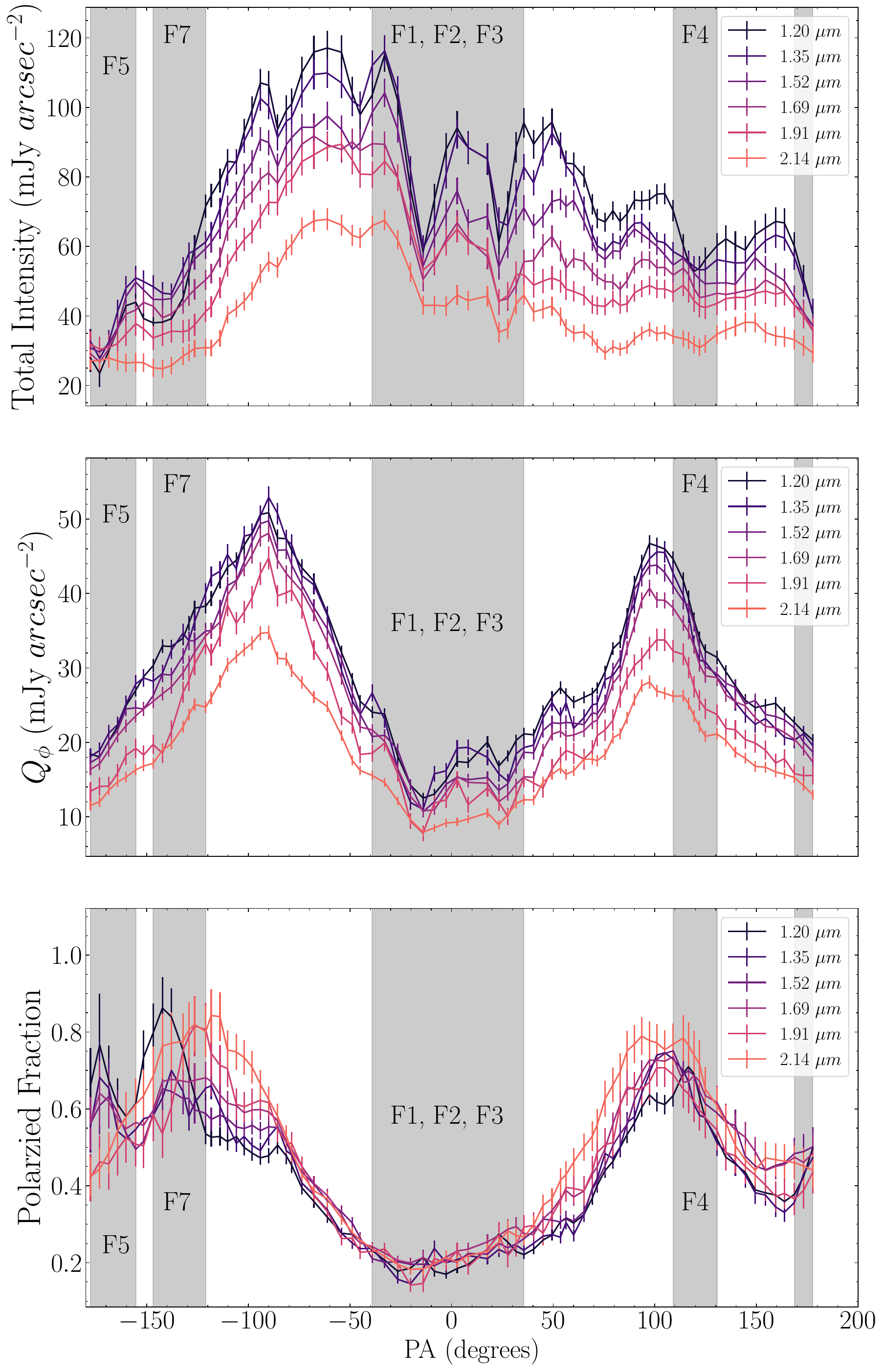}
    \caption{The azimuthal surface brightness profiles in total intensity (top panel) and $Q_{\phi}$ (middle panel), and $Q_{\phi}$/total intensity (bottom panel), measured along the best-fit ellipse. The PAs refer to position angles relative to the celestial north. The shaded regions correspond to the approximate PAs covered by the labeled features in Figure \ref{fig: RGB} except F6, as F6 is not part of the outer disk.}
    \label{fig:az_profile}
\end{figure}

Figure \ref{fig:az_profile} shows the azimuthal surface brightness profiles in total intensity (top panel), $Q_{\phi}$ (middle panel) and polarized fraction (bottom panel).  The different colors correspond to the wavelength bins described in Section \ref{sec: imaging analysis}. The shaded regions correspond to the disk features described in Section \ref{subsec: disk features}. The total intensity is measured from the dataset described in Section \ref{subsec: ti data}; this is the same dataset presented in \cite{Uyama_2020}, but processed with a different RDI algorithm described in Section \ref{subsec: RDI PDI} that tries to minimize over-subtraction of the disk signal. Our total intensity scattering profiles share the same overall shapes as those in \cite{Uyama_2020}, showing higher brightness in the forward scattering direction (north side) and a sharp dip at the north side cavity of the disk. However, our profiles show a bright portion at PA$\sim -20\degree$ to $-80 \degree$, as opposed to the more extended ``dip'' region in \cite{Uyama_2020}. This is likely due to the improved RDI reducing over-subtraction, as the brightness now goes back up to the model level beyond the cavity without a slightly extended dip shown in \cite{Uyama_2020}. The surface brightness monotonically decreases as a function of wavelength for both total intensity and polarized intensity. From J band to K band, the drop in brightness is about a factor of two, while the degrees of polarization remain similar. This trend is consistent with the results in \cite{Uyama_2020}, but differs from GPI data where the H band is brighter in both total and polarized intensity \citep{Monnier_2019}. This is not too surprising considering the heavy dependence of peak brightness on the Strehl ratio. As a result, the polarized fraction is a better quantity for comparison. The polarized fraction as a function of PA (bottom panel of Figure \ref{fig:az_profile}) shows a similar shape as the polarized intensity, and is slightly higher at shorter wavelengths for the entire disk except in the regions with darkening features or the artifact. We do see agreement in wavelength dependence of the polarized fraction with the results in \citep{Monnier_2019}. The peak polarized fraction usually appears near a scattering angle of $90\degree$ \citep{Benisty_2022} which should align with the semi-major axis under the assumption of a simple inclined circular disk. Our data suggests a different PA of the disk than the one inferred from ellipse fitting. \cite{Monnier_2019} also found that the PA of the axis of peak polarized fraction is more indicative of the true PA of the disk based on radiative transfer modeling. We describe the PA inferred from the polarized fraction in more detail in Section \ref{sec: modeling}. 

\subsection{Scattering Phase Functions} 
\label{subsec: data phase functions}
We measure the total intensity, polarized intensity and polarized fraction as a function of scattering angle using \diskmap \citep{Stolker_2016_diskmap}. These correspond to the apparent scattering phase functions of the disk. The scattering angle is defined as the angle between the incident light (from the star) and scattered light (towards the observer) in the scattering plane, implicitly assuming an optically thin disk. A scattering angle $<90\degree$ would represent forward scattering, while $>90\degree$ represents backward scattering. For a specified geometry of the scattering surface, \diskmap calculates the deprojected radius and scattering angle at every pixel coordinate of the disk. The observed brightness is then corrected by a factor of $r_{d}^{2}$, where $r_{d}$ is the deprojected radius of a point on the scattering surface. Then, the scattering angles are binned into 60 bins over the available range of scattering angles. The east and the west half of the disk would roughly cover the same range of scattering angles. We do not treat them separately. This means that each scattering angle bin often consists of pixels from both the east and west side of the disk. The average intensity of the pixels in each bin and their standard deviation make up a data point and its error. For HD 34700, we measure these intensities over a narrow strip over the bright dominant ring. The strip is centered at the ellipse which is fit to the brightest part of the ring and is 6 pixels in width. We also mask the north side cavity from $\text{PA}=60\degree$ to $120 \degree$ for our measurements. We use the surface scale height of a flared disk: 
\begin{align}
    \label{eqn: scale height}
    h(r) = h_0(\frac{r}{R_0})^{\beta}
\end{align}
where $R_0=172.4$ au is the reference radius defined by the best-fit semi-major axis, $h_0=17$ au is the disk scale height at the reference radius, and $\beta=1.125$ is the disk flaring index. The $h_0$ and $\beta$ values are the best-fit disk model in \cite{Monnier_2019}. For the inclination, we use the inferred value from the best-fit ellipse from this work of $41\degree$. For the PA, we use $92\degree$ which corresponding to our axis of maximum polarized intensity, instead of the $74\dotdeg 3$ from our ellipse fitting. We discuss further in Section \ref{subsec: model results} how the models compare to the data and why the models favor the PA corresponding to the axis of maximum polarization instead of the semi-major axis of the best-fit ellipse. Given these parameters, the scattering angle that we can probe for the disk ranges from $\sim 40\degree$ to $\sim 125\degree$.

For single scattering events on a collection of particles, the intrinsic scattering phase functions represent the probability densities of scattering occurring at every angle, and would correspond to the scattering matrix elements \citep{Bohren_Huffman_1998}. However, for an optically thick disk, multiple scattering and limb brightening effects make the apparent scattering phase functions deviate from the intrinsic ones \citep{Tazaki_2023}. Because of this, we are measuring the apparent scattering phase functions using \diskmap. We cannot measure the intrinsic ones without knowing the ratio of single scattering vs multiple scattering that we see at every point in the disk together with the angles that multiply scattered photons arrived from. But we can account for the effect of multiple scattering and limb brightening when simulating scattered disk images using Monte Carlo radiative transfer modeling. By measuring the apparent scattering phase functions of both the data and model images, we can compare them to gain insights on dust grain properties. We describe this modeling process further in Section \ref{sec: modeling}. For simplicity, we will use the term ``scattering phase functions'' to refer to the apparent ones henceforth.

\begin{figure}
    \centering
    \includegraphics[width=0.48\textwidth]{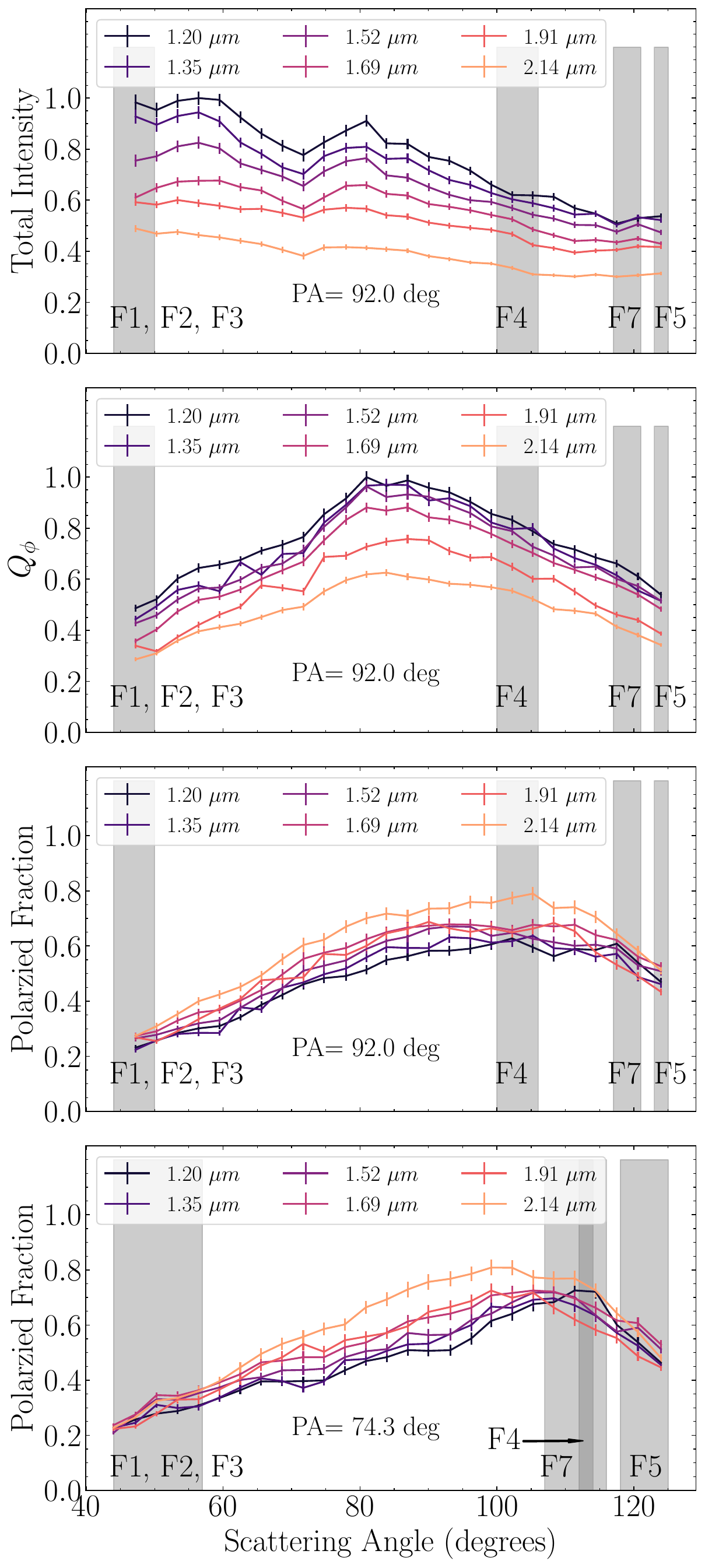}
    \caption{The normalized apparent scattering phase functions for total intensity $I$ and azimuthal Stokes $Q_{\phi}$ are shown in the top and second panels, respectively. The scattering angle is defined as the angle between the incident light from the star and the scattered light to the observer, with forward scattering having angles $<90 \degree$. We show two polarized fraction phase functions measured assuming different PAs in the third and fourth panels. PA=$92\degree$ corresponds to the PA of the axis of maximum polarization, and PA=$74\dotdeg 3$ corresponds to the PA of the semi-major axis of the best-fit ellipse described in Section \ref{subsec: ring}. The shaded regions indicate the approximate scattering angles covered by the labeled features in Figure \ref{fig: RGB} except F6, which is not part of the outer disk.}
    \label{fig: SPFs}
\end{figure}

Figure \ref{fig: SPFs} shows the final results of our phase function estimation. The top two panels show the measurements for irradiation-corrected total intensity $I$ and azimuthal Stokes $Q_{\phi}$, both normalized such that the peak scattering probability is 1. This is purely for aesthetic purposes. As we do not have access to all scattering angles, there is no way to properly normalize the integrated probability to 1. The bottom two panels both show the polarization fraction ($Q_{\phi} / I$) as a function of scattering angle, but assume different PAs of the disk relative to celestial north when defining the PA of each pixel coordinate. The third panel assumes a disk PA of $92\degree$, corresponding to the PA of the axis of maximum polarized fraction. The last panel assumes a PA of $74\dotdeg 3$, corresponding to the PA of the semi-major axis of the best-fit ellipse to the disk ring's peak brightness. The peak polarized fraction occurs at larger scattering angles when ${\rm PA}=74\dotdeg 3$. We discuss how our models compare to the data at these two PAs in Section \ref{subsec: model results}. 

The inner mis-aligned disk can cast shadows that produce features and alter the brightness of the outer disk \citep{Brauer_2019_GGTauri}. Without knowledge of the inner disk properties and an estimate of its influence on the radiative transfer, we can only make limited observations of the total and polarized intensity phase functions. On the other hand, the polarized fraction primarily depends on the scattering properties of the local dust grains responsible for the scattering and is very weakly dependent on other common disk model parameters (e.g. dust mass and scale height; Dykes et al. 2024 submitted). Therefore, compared to the intensity phase functions, the polarized fraction is more informative of the properties of the dust grain population for our case.

The total intensity phase function is expected to favor forward scattering (small scattering angles), drop steeply in the first $\sim 30\degree$, and have either a plateau or a slightly rising trend towards larger angles \cite{Benisty_2022, Min_2016}. Due to the limited range of scattering angles we are probing, our total intensity measurement only captures the plateau part of the phase function. 
We see a high degree of polarization, with maximum polarized fractions of around 0.6-0.8 across J, H, and K bands. We also see that the polarized fractions are higher at longer wavelengths, peaking at $\sim$0.8 in K band, $\sim 0.7$ in H band, and $\sim$0.6 in J band. We discuss in more detail how our models match these observations of the data in Section \ref{sec: modeling}. 

\section{Disk Modeling with \MCFOST}\label{sec: modeling}
In this section, we carry out modeling of the outer disk, resolved in CHARIS images in scattered light, at the wavelengths corresponding to the data wavelength bin centers. We use the 3D Monte Carlo radiative transfer code with ray-tracing, \MCFOST \, \citep{Pinte_2006, Pinte_2009, Pinte_2022}, to produce modeled images in scattered light.

\MCFOST computes the optical properties of dust using Mie theory, typically treating the grains as homogeneous spheres \citep{Pinte_2006, Bohren_Huffman_1998}. It also has the capability to add in porosity, volume mixing of different components, or to change the grain type to a distribution of hollow spheres (DHS). An effective optical index of the grains is computed using the Bruggeman rule before any Mie theory computation.

Compared to more exact methods such as the superposition T-matrix \citep{Mackowski_1996_Tmatrix} or the discrete dipole approximation (DDA; \citealt{Purcell_1973_DDA, Draine_1994_DDA}), Mie theory has limitations in how accurately it can produce scattered light images using its approximation of optical properties of dust grains \cite{Min_2016}. But in exchange, the reduced computational cost allows us to explore a parameter grid with more disk parameters. 

The inner disk is only partially detected in CHARIS images and we cannot infer its properties from our data. While the inner disk contributes a negligible amount ($\sim 1\%$) in thermal emission compared to the starlight at our observing wavelengths ($1\mu m -2.1\mu m$), we cannot account for the effects of shadowing and scattered star light by the inner disk. Because of this, we focus on examining the polarized fraction of our data and models. The degree of polarization is insensitive to shadowing effects; in even more morphologically complex disks than HD 34700 A's (e.g. AB Aurigae), the degree of polarization is also only very weakly dependent on other common disk model parameters (e.g. dust mass and scale height; Dykes et al. 2024 submitted). Therefore, the degree of polarization primarily depends on the dust grains responsible for the scattered light, and Mie theory simulations still offer us insights on dust properties \citep{Min_2005, Mulders_2013, Arriaga_2020}. In addition, an examination of the degree of polarization at finer wavelength resolutions was not possible for J, H, and K broadband datasets in the past, and is now made possible with the addition of our \CHARIS PDI dataset combined with a previous \CHARIS total intensity dataset, which we binned into 6 wavelengths from $1.2 \mu m$ to $2.1 \mu m$. In this section, we carry out modeling under Mie theory of the outer disk resolved in our data and discuss the results and their implications. 

\begin{table*}
\caption{Model Parameter Grid}
\label{table: model grid}
\begin{tabular}{cccccccccc}
\hline
\multicolumn{5}{c}{Fixed Parameters} & \multicolumn{3}{c}{Value} & {References}\\
\hline
\multicolumn{5}{c}{$\Rstar$ (Aa, Ab)} & \multicolumn{3}{c}{3.46, 3.4 $\Rsun$} & 1\\

\multicolumn{5}{c}{$T_{eff}$ (Aa, Ab)} & \multicolumn{3}{c}{5900, 5800 K} & 1\\

\multicolumn{5}{c}{$\Mstar$ (Aa, Ab)} & \multicolumn{3}{c}{2.0, 2.0 $\Msun$} & 1\\

\multicolumn{5}{c}{Distance} & 
\multicolumn{3}{c}{350.5 pc} & 2\\

\multicolumn{5}{c}{Inclination, $i$} & \multicolumn{3}{c}{$41\dotdeg 0$}\\

\multicolumn{5}{c}{Dust Mass} & 
\multicolumn{3}{c}{$1.2\times 10^{-4}$ $\Msun$} & 1\\

\multicolumn{5}{c}{Surface density exponent, $\epsilon$} & 
\multicolumn{3}{c}{1.0} & \\

\multicolumn{5}{c}{Reference scale height, $h_0$} & 
\multicolumn{3}{c}{17 au} & 1\\

\multicolumn{5}{c}{Reference radius, $R_0$} & 
\multicolumn{3}{c}{175 au} & 1\\

\multicolumn{5}{c}{Disk inner radius, $R_{in}$} & 
\multicolumn{3}{c}{$172.4$ au} & \\

\multicolumn{5}{c}{Disk outer radius, $R_{out}$} & 
\multicolumn{3}{c}{400 au} & \\

\multicolumn{5}{c}{Minimum grain size, $a_{min}$} & 
\multicolumn{3}{c}{0.1 $\mu m$} & \\

\multicolumn{5}{c}{Maximum grain size, $a_{max}$} & 
\multicolumn{3}{c}{1000 $\mu m$}\\

\hline
\multicolumn{4}{c}{Varied Parameters} & \multicolumn{2}{c}{Range} & \multicolumn{2}{c}{\# Points} & Spacing \\
\hline

\multicolumn{4}{c}{Disk flaring index, $\beta$} & \multicolumn{2}{c}{[1.0, 1.5]} & \multicolumn{2}{c}{4} & Linear\\

\multicolumn{4}{c}{Small grain mass fraction} & \multicolumn{2}{c}{[0.2, 0.8]} & \multicolumn{2}{c}{5} & Linear\\

\multicolumn{4}{c}{Transition grain size$^{*}$, $a_{t}$} & 
\multicolumn{2}{c}{[0.25, 2.0] $\mu m$} & \multicolumn{2}{c}{4} & Log\\

\multicolumn{4}{c}{Grain size power-law, $p$} & \multicolumn{2}{c}{[-3.2, -3.8]} & \multicolumn{2}{c}{3} & Linear\\

\multicolumn{4}{c}{Dust grain porosity} & \multicolumn{2}{c}{[0.2, 0.8]} & \multicolumn{2}{c}{4} & Linear\\

\multicolumn{4}{c}{Position Angle} & \multicolumn{2}{c}{[$74\dotdeg 3$, $92\degree$]} & \multicolumn{2}{c}{2} & Choice\\

\multicolumn{4}{c}{Grain type} & \multicolumn{2}{c}{[Mie, DHS]} & \multicolumn{2}{c}{2} & Choice\\

\multicolumn{4}{c}{Grain composition} & \multicolumn{2}{c}{[Silicates, AMC]} & \multicolumn{2}{c}{2} & Choice\\

\hline
\end{tabular}\\
{* }{The transition grain size is the maximum size of the small grain population and also the minimum size of the large grain population. The two populations share the same grain size power-law distribution, $a_{exp}$}\\
{References: 1. \cite{Monnier_2019}; 2. \cite{Gaia_2021}.}
\end{table*}

\subsection{Model Parameters and Modeling Strategy}
\label{subsec: model grid}

We adopt the stellar properties of the central binary from \cite{Monnier_2019}, including stellar radii, effective temperatures, and masses. We update the distance of the system according to the \Gaia EDR3 measurement: $350.5^{+2.5}_{-2.4}~ {\rm pc}$ \citep{Gaia_2021}. We model the disk as a circular, single-component, axisymmetric, flared disk around the central binary, observed at a non-zero inclination. The disk's surface density is defined as:
\begin{align}
    \label{eqn: surface density}
    \Sigma(r) \propto 
    \begin{cases}
        \ r^{-\epsilon},\;\; &R_{\rm in} \le r \le R_{\rm out}\\
        e^{-\frac{(R_{\rm in}-r)^2}{2}},\;\; &r<R_{\rm in}\\
        e^{-\frac{(r-R_{\rm out})^2}{2}},\;\; &r>R_{\rm out}
    \end{cases}
\end{align}
Where $R_{\rm in}$ and $R_{\rm out}$ are the radii of the inner and outer edges of the disk. The scattering scale height is defined by Equation \ref{eqn: scale height} in Section \ref{subsec: ring}. And finally, the number density of the dust population is defined as:
\begin{align}
    \label{eqn: number density}
    N(a) \propto a^{p}
\end{align}
where $a$ is the radius of the dust grain, and $p$ is the grain size power-law index \citep{Dohnanyi_1969, Mathis_1977}.

Due to the aforementioned limitations of Mie theory and simplifications in the disk model, a conventional $\chi^{2}$ goodness-of-fit metric of the model image to the data image would not be a reliable way to evaluate the model and does not provide further information beyond that gained from qualitative comparisons. Therefore, we simply select by eye the models that best match the data qualitatively, excluding the cavity region on the north side and the darkening features. This essentially requires the scattering profiles of the models to have shapes resembling those of the data, while leaving a generous room for the unaccounted effects. Then, we consider models with polarized fractions within $\sim 20\%$ across all scattering angles to be an acceptable fit. Due to the qualitative nature of the comparison, instead of finding a best-fit model, we present in section \ref{subsec: model results} one representative model that is the best qualitative match to the data. Alongside the qualitative match, we also show and discuss models that are qualitatively different, what parameters led to these qualitative changes and what the implications are on dust grain properties. 

We carry out the modeling by constructing a coarse parameter grid for the disk parameters, as a high-dimensional fine grid is prohibitively expensive in computation time. We fix the parameters that are constrained by other independent observations: stellar properties and dust mass are fixed at the values in \cite{Monnier_2019}. Distance is fixed at the \Gaia EDR3 mean value, 350.5 pc \citep{Gaia_2021}. We also fix parameters that can be well constrained by imaging analysis alone, such as the inner radius, $R_{\rm in}$, and inclination, $i$. Finally, to reduce the number of dimensions and computational cost, some parameters are fixed at their canonical or approximate values after testing that they negligibly influence the simulated images within a physically reasonable range.  These fixed parameters include the surface density exponent, $\epsilon$, and outer radius, $R_{\rm out}$. Our final parameter grid is shown in Table \ref{table: model grid}. We use two dust populations for our models following previous efforts by \cite{Monnier_2019, Uyama_2020}, one for small grains and one for large grains. We label the minimum and maximum grain sizes $a_{\rm min}$ and $a_{\rm max}$, and the transition size between the two populations $a_{t}$. Both populations follow the same number density distribution given by Equation \ref{eqn: number density} with the same $p$ (i.e. the two grain populations share the same number density slope, with a discontinuous transition at $a_{t}$). For each model, the two populations have a fixed mass fraction. This means that for each model, a fixed amount of dust mass will be distributed between $a_{\rm min}$ and $a_{t}$, while the remaining dust mass will be distributed between $a_{t}$ and $a_{\rm max}$. We simulate both Mie  \cite{Bohren_Huffman_1998} and DHS \cite{Min_2003_DHS, Min_2005} grains, and include two different compositions in our models: astronomical silicates \cite{Draine_Lee_1984} and an amorphous carbon mixture following the prescription given by \cite{Min_2011} with a carbon partition parameter $\omega=1$.

\begin{figure*}
    \centering
    \includegraphics[width=\linewidth]{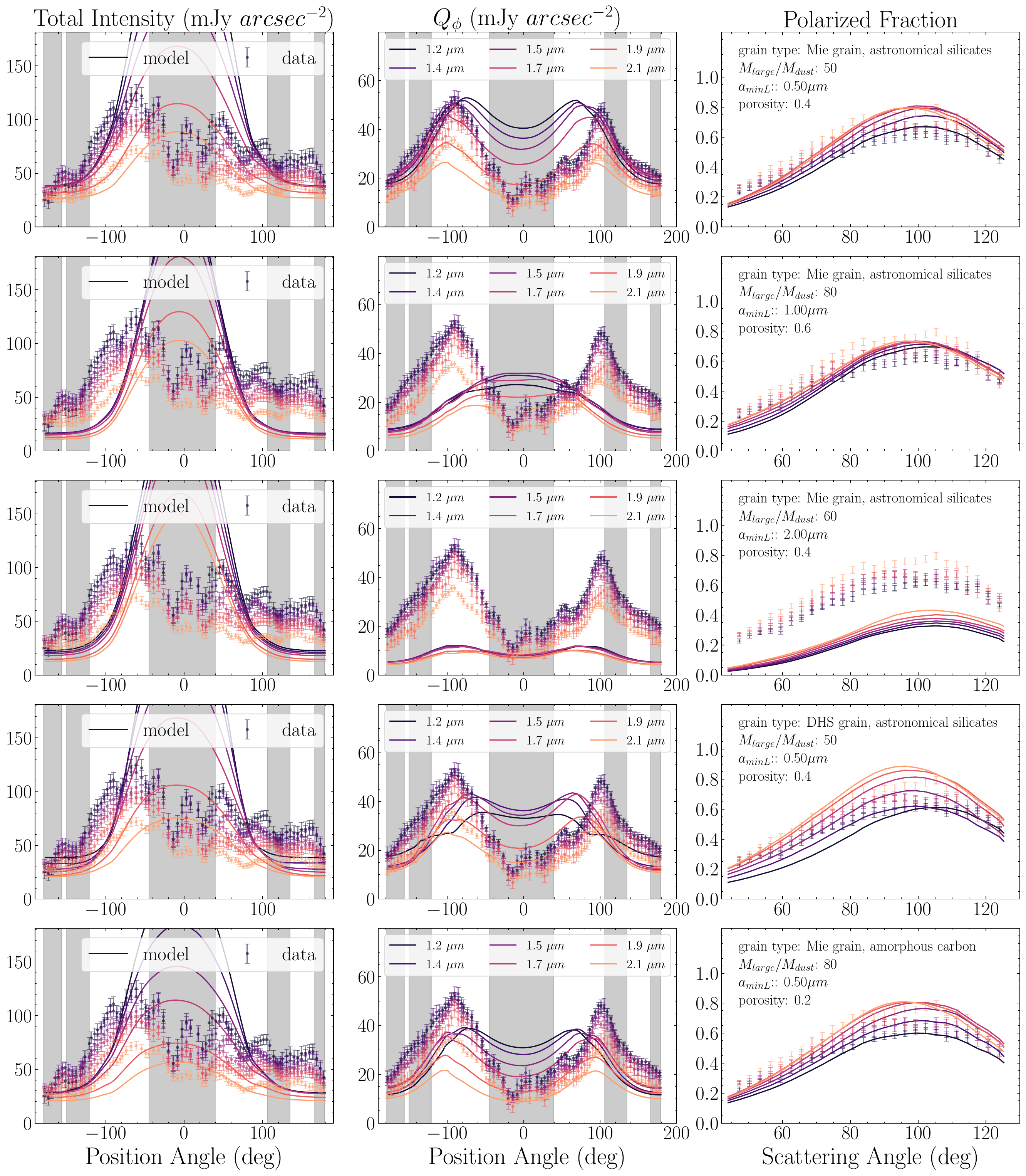}
    \caption{Comparison of models to the data. The azimuthal surface brightness profiles in total intensity and $Q_{\phi}$ are shown in the left and middle columns, respectively. The polarized fraction as a function of scattering angle is shown in the right column. The solid lines are the measurements for the PSF-convolved model images, and the data points with error bars are measurements for our datasets using the scattering surface geometry of the corresponding models. Each row shows a different set of parameter models from the model grid, with key parameters annotated in the right column plots. We find no model that can simultaneously fit the total intensity and polarized intensity, which is not surprising considering the significantly simplified model compared to the feature-rich disk, in addition to the effect of the inner disk that we ignored. However, as the polarized fraction reflects the local properties of the dust particles responsible for the scattering, we are able to find models that fit the polarized fraction and they suggest significant masses in sub-micron as well as micron sized grains.}
    \label{fig: model results}
\end{figure*}

\subsection{Model Results}
\label{subsec: model results}
We find that we cannot identify a model that fits well (within a few percent in regions without disk features) to all three quantities: 1. total intensity; 2. polarized intensity; 3. polarized fraction. However, this is not surprising given the limitations we have discussed in Section \ref{sec: modeling}. We summarize the qualitative influence of the model parameters on the scattering profiles in this section. 

For every set of model parameters in the parameter grid, \MCFOST outputs model images in total and polarized intensity. We convolve each model with the instrumental PSF, and then measure the surface brightness profiles and scattering phase functions following the same methods in Section \ref{subsec: SB profiles} and Section \ref{subsec: data phase functions}, respectively. Because the flaring index of the disk surface and the PA for each model can vary, the geometry of the scattering surface can change for each model. Since the scattering phase functions presented in Section \ref{subsec: data phase functions} assume a specific scattering surface scale height and PA, they cannot be directly compared to those measured from the model images, which have different scattering surfaces. Therefore, to properly compare the scattering phase functions between each model and our data, we re-measure the scattering phase functions for the data using the scattering geometry of each model. We find that the models can closely reproduce the polarized fraction phase function of the data when assuming ${\rm PA}=92\degree$, for which the polarized fraction peaks between scattering angles of $95\degree-110\degree$. If we use ${\rm PA}=74\dotdeg 3$, the polarized fraction of the data skews towards large scattering angles and peak around $100\degree - 115\degree$ as shown in Figure \ref{fig: SPFs}, which is a slightly worse fit to the models. Because the polarized fractions from radiative transfer models are grounded in scattering theory and are determined by the scattering phase functions of the dust population, it is more robust than the PA inferred from a simple ellipse fit to the disk ring's peak brightness, which cannot capture the complexities of the various disk features. This agrees with the conclusion in \cite{Monnier_2019} that the axis of peak polarization is more indicative of the disk's true PA, as opposed to the PA of the semi-major axis of the best-fit ellipse. 
In Figure \ref{fig: model results}, we show a set of models with distinct qualitative features in their scattering profiles. We find that the number density exponent had a small effect on the models within the range [3.2, 3.8], and its effect is degenerate with changing the mass fraction and $a_{minL}$ to produce a dust population with similar effective optical indices. Therefore, in Figure \ref{fig: model results}, we choose to present models with a number density parameter of 3.5 \citep{Dohnanyi_1969, Mathis_1977} among the ones with similar qualitative features. The top-row panel shows a model that is the best qualitative match to the data, which still cannot fit the total and polarized intensity very well. The subsequent rows show models that have distinct qualitative features different from the data, which inform us about one or multiple physical properties of the dust population of the disk.

\subsubsection{Qualitative Fit}\label{subsec: qualitative fit}
The top row of Figure \ref{fig: model results} shows a model that qualitatively best matches the data by eye, which still cannot match the total and polarized intensity of the data well. This is not surprising given the simplified model compared to the feature-rich disk, in addition to the effect of the inner disk that we ignored. However, as the polarized fraction reflects the local properties of the dust particles responsible for the scattering, it still provides meaningful insights on the grain properties. Indeed, the polarized fraction of the model and data match in terms of the overall degree of polarization, the wavelength dependence, as well as the scattering angle dependence. We find that significant mass fractions in both sub-micron sized grains and micron-sized grains are required to reproduce the polarized fraction as a function of scattering angle. As a result, a single population with a constant number density power-law cannot reproduce the shape and the level of surface brightness seen in the data. The small grains are responsible for more isotropic Rayleigh scattering and are needed to produce enough brightness at larger scattering angles. 

\subsubsection{Models with Insufficient Sub-micron Dust Grains}\label{subsec: less small grain}
The second row of Figure \ref{fig: model results} shows a model with much less mass in the small grain population. While we can still match the polarized fraction with a higher porosity, we see that the lack of small grains produces much less polarized light at larger PAs corresponding to $90\degree +$ scattering angles. The large grains are responsible for skewing the surface brightness towards extreme forward scattering and insufficient brightness for the south side.

\subsubsection{Models with Insufficient Micron Sized Grains}\label{subsec: 2 micron divide}
The third row of Figure \ref{fig: model results} shows a model where the dividing line between small and large grains is at $2\mu m$. Similar to the model shown in the second row, the excess in forward scattering remains a problem. In addition, with a number density power-law of $-3.5$, this leads to a much smaller fraction of grains with sizes comparable to the observing wavelengths $\sim 0.5-2.0\mu m$. Grains much larger than our observing wavelengths are less efficient at scattering incident light, resulting in much lower surface brightness, as well as a lower polarized fraction. 

The first three rows show that it is necessary to have significant mass fractions of grain sizes that are sub-microns and microns to simultaneously produce the high degree of polarization, bright peaks at east and west sides of the disk, and sufficient overall brightness in total intensity and polarized intensity. It is difficult to explain the physical processes that can generate this double population with a discontinuous transition at the grain size $a_{\rm min}$. It has been found in laboratory experiments that simulations using Mie/DHS particles can suggest a sub-micron grain population that is not found in the measured dust distribution in the lab \citep{Min_2005}. It is possible this reflects the rough surfaces with sub-micron-scale structures of real, irregular dust grains \citep{Min_2005}. 

\subsubsection{Models with Different Grain Types and Compositions}\label{subsec: DHS and AMC}
The fourth and fifth rows of Figure \ref{fig: model results} show models using DHS grains and Mie grains with a carbon-rich composition, respectively. Within the model parameter space we explored, we find that these two grain types/compositions can produce matching polarized fractions, but struggle with forward scattering excess more than Mie grain models and produce less scattered light in general. The silicate DHS grain models otherwise share similar results as the silicate Mie grain models. The Mie grain model with amorphous carbon generates a higher degree of polarization, such that fewer small grains and less porosity are needed to match the polarized fraction. Given the limitations of our models and the small parameter space we are able to explore, the relatively worse fits to the total and polarized intensities do not offer strong arguments against DHS grains or grains with amorphous carbon. Future work  with more quantitative modeling and access to potentially fully resolved images of both the inner and outer disk may be able to provide more insights regarding grain types and composition.

\section{Summary and Discussion} \label{sec: discussion}
We have presented in this paper spectro-polarimetry of the transition disk around HD~34700 A obtained using Subaru/\SCExAO+\CHARIS broadband (J, H, and K). We fit a simple ellipse to the disk ring using two different methods; assuming a circular disk, we obtain an inclination of $41\degree \pm 1\dotdeg 1$ and a PA of $74\dotdeg 3 \pm 1\dotdeg 2$. 
Combining our polarized light observation with previous observations in total intensity \citep{Uyama_2020}, and leveraging the unique capability of CHARIS to simultaneously take polarimetry data at all wavelength channels (22 wavelengths across JHK), we examine the surface brightness and polarized fraction as a function of wavelength and scattering angle. We see that shorter wavelengths are brighter in both total intensity and polarized intensity, where the brightness drops by a factor of $\sim 2$ from $1.2\mu m$ to $2.1\mu m$. The degree of polarization is higher at longer wavelengths, where the polarized fraction peaks around 0.6 in J band, 0.7 in H band, and 0.8 in K band. The maximum degrees of polarization in J and H are both higher compared to \cite{Monnier_2019}, where they found peak fractions of around 0.5 and 0.6 in J and H bands, respectively. Compared to other planet-forming disks, HD~34700A already has one of the highest degrees of polarization based on the measurements in \cite{Monnier_2019}, as compiled in \cite{Tazaki_2022}. Our measurements would shift it slightly higher, bringing HD~34700A to a very similar level to the disk, UX Tau A \citep{Tanni_2012}. We also detect an inner arc in our data, which is part of the mis-aligned inner disk of the system \cite{Columba_2024}. 

To understand what the observations imply about the dust population in the outer disk of HD~34700 A, we carry out radiative transfer modeling using the 3D Monte Carlo radiative transfer code, \MCFOST \citep{Pinte_2006}. We make only qualitative comparisons between our models and the data, due to several limitations: 1. We lack knowledge of the inner disk properties, which could cast shadows and re-scatter light from the stars onto the outer disk, and alter its surface brightness. 2. Optical properties of dust grains are calculated using Mie theory instead of more exact methods that are more expensive for producing imaging simulations at many wavelengths. 3. The disk is simplified as a circular, single-component, axisymmetric ring with a flared surface. Our models struggle to produce a close match to the total and polarized intensity, which is not surprising given the aforementioned limitations. Even for optically thin disks, Mie theory calculations following Bruggeman mixing rules have been shown to struggle to simultaneously match total intensity and polarized fraction \cite{Arriaga_2020}. Instead, we focus our comparison on the polarized fraction as a function of scattering angle, which only depends on the scattering properties of the dust grain populations and the scattering geometry. We find that a two-population distribution with a broken number density power-law is necessary to reproduce the surface brightness of the data, where significant mass fractions in both sub-micron grains and micron-sized large grains are required. Such a distribution appears somewhat unphysical. This could be because Mie models are using an extra population of small grains to simulate the effects of sub-micron structures and monomers of large dust aggregates. This is consistent with the findings in \cite{Tazaki_2022}, where they investigated the degree of polarization of more complex dust aggregates. They found that micron-sized dust aggregates consisting of 0.1$\mu m$ and 0.2$\mu m$ monomers can have polarized fractions beyond 0.6 at near-infrared wavelengths, consistent with our models that contain a large population of sub-micron dust particles.

We also run models with DHS grains and Mie grains containing amorphous carbon. While these two grain types struggle a bit more to produce the surface brightness profiles, we also find models that match the polarized fraction of the data within the parameter space we explored. We find that DHS particles share similar results as Mie particles, and interestingly, grains containing amorphous carbon require much less porosity to produce a high level of degree of polarization. 

For future work, it would be intriguing to employ more detailed grain scattering models of individual monomers and larger aggregates (e.g., \citealt{Purcell_1973_DDA, Draine_1994_DDA, Mackowski_1996_Tmatrix, Tazaki_2022, Tazaki_2023}). This would allow for a deeper investigation into whether realistic grains with rough surfaces and intricate aggregate structures can effectively explain the observed broken power-law distribution and provide a better constraint on the grain sizes. Another welcome improvement to the disk model would be to include the inner disk. This would be possible, for example, at wavelengths corresponding to H$_\alpha$ filters ($\sim 0.6 \mu m$) where the inner disk has been fully resolved \citep{Columba_2024}. Finally, an excess in the spectral energy distribution (SED) was observed at $\sim 3-10 \mu m$, which suggested a potential warm disk component very close to the central binary \citep{Monnier_2019}. Therefore, an additional source of incident light that accounts for such a component to match the SED would be another welcome improvement. 

\label{lastpage}

\section*{acknowledgments}
This research is based in part on data collected at the Subaru Telescope, which is operated by the National Astronomical Observatory of Japan. We are honored and grateful for the opportunity of observing the Universe from Maunakea, which has the cultural, historical, and natural significance in Hawaii.  T.D.B.~gratefully  acknowledges support from the National Aeronautics and Space Administration (NASA) under grant \#80NSSC18K0439 and from the Alfred P.~Sloan Foundation. 

This research is also based upon work supported by the National Science Foundation Graduate Research Fellowship under Grant No. 2021-25 DGE-2034835 for author B. Lewis. 

Any opinions, findings, and conclusions or recommendations expressed in this material are those of the authors(s) and do not necessarily reflect the views of the National Science Foundation. 

Ryo Tazaki acknowledges funding from the European Research Council (ERC) under the European Union's Horizon Europe research and innovation program (grant agreement No. 101053020, project Dust2Planets).

Software: \MCFOST \citep{Pinte_2006, Pinte_2009, Pinte_2022}, \diskmap \citep{Stolker_2016_diskmap}, \texttt{NumPy} \citep{numpy}, \texttt{IPython} \citep{ipython}, \texttt{Jupyter Notebooks} \citep{jupyter}, \texttt{Matplotlib} \citep{matplotlib},  \texttt{Astropy} \citep{astropy:2013, astropy:2018}, \texttt{SciPy} \citep{scipy}

\section*{Data Availability Statement}
The data underlying this article are available at \url{https://stars.naoj.org/}. The datasets underlying this article are reduced using the open-source python library: the \CHARIS Data Reduction Pipeline (\CHARIS DRP) \citep{Brandt_2017_CHARISPipeline} at \url{https://github.com/PrincetonUniversity/charis-dep}

\bibliography{charis}
\bibliographystyle{apj_eprint}

\end{document}